# Interstellar Detection of Methyl Isocyanate $CH_3NCO$

# in Sgr B2(N):

# A Link from Molecular Clouds to Comets

D. T. Halfen[1], V. V. Ilyushin[2], and L. M. Ziurys[1]

[1] Department of Chemistry and Biochemistry, Department of Astronomy, Arizona Radio Observatory, and Steward Observatory, University of Arizona, Tucson, AZ 85721; email: halfendt@as.arizona.edu

[2] Institute of Radio Astronomy of the National Academy of Sciences of Ukraine, Chervonopraporna 4, 61002, Kharkov, Ukraine

**Abstract**

A new interstellar molecule, $CH_3NCO$ (methyl isocyanate), has been detected using the 12 m telescope of the Arizona Radio Observatory (ARO). $CH_3NCO$ was identified in spectra covering 48 GHz (68–116 GHz) in the 3 mm segment of a broadband survey of Sgr B2(N). Thirty very favorable rotational lines ($K_a = 0$ and $K_a = 1$ only; $E_u < 60$ K) originating in five consecutive transitions ($J = 8 \rightarrow 7, 9 \rightarrow 8, 10 \rightarrow 9, 11 \rightarrow 10$, and $12 \rightarrow 11$) in both the A and E internal rotation species are present in this frequency range. Emission was observed at all of the predicted frequencies, with seventeen lines appearing as distinct, uncontaminated spectral features, clearly showing the classic a-type, asymmetric top pattern, with $T_R^* \approx 20$–70 mK. The $CH_3NCO$ spectra also appear to exhibit two velocities components near $V_{LSR} \approx 62$ and 73 km s$^{-1}$, both with $\Delta V_{1/2} \approx 10$ km s$^{-1}$ - typical of molecules such as $CH_2CHCN$, HNCO, and $HCOOCH_3$ in Sgr B2(N). The column density of $CH_3NCO$ in Sgr B2(N) was determined to be $N_{tot} \approx 2.3 \times 10^{13}$ cm$^{-2}$ and $1.5 \times 10^{13}$ cm$^{-2}$ for the 62 and 73 km s$^{-1}$ components, corresponding to fractional abundances, relative to $H_2$, of $f \approx 7.6 \times 10^{-12}$ and $5.0 \times 10^{-12}$, respectively. $CH_3NCO$ was recently detected in volatized material from comet 67P/Churyumov-Gerasimenko by Rosetta's Philae lander, with an abundance ~1.3% of water; in Sgr B2(N), $CH_3NCO$ is roughly ~0.04% of the $H_2O$ abundance.

## 1. Introduction

Comets are thought to contain relatively pristine material from the origin of the solar system, having condensed directly out of the pre-solar nebula (e.g. Mumma & Charnley 2011). It is postulated that comets may have even delivered some of the water and organic matter found on the Earth via impacts (e.g. Hartogh et al. 2011). Over twenty-two molecules have been identified in comets via radio observations (Crovisier et al. 2004), including organic species such as formamide and ethylene glycol (Biver et al. 2014).

Recently, the Cometary Sampling and Composition (COSAC) mass spectrometer aboard the Rosetta spacecraft's Philae lander detected methyl isocyanate $CH_3NCO$ in the comet 67P/Churyumov-Gerasimenko (Goesmann et al. 2015). This compound, as well as other organic molecules, was found in material evaporated from the comet's surface after the initial touchdown of the lander, which rebounded off the object twice before a final landing. Mass spectra taken before the initial landing and after the final touchdown were void of these species, suggesting they were likely evaporated by the initial lander impact (Goesmann et al. 2015). $CH_3NCO$ is the methyl derivative of isocyanic acid HNCO, which was also detected. This discovery inspired a closer look for $CH_3NCO$ in our spectral-line survey of Sgr B2(N). Note that Pulliam et al. (2013) reported a detection of this molecule in a conference poster, but no publication appeared thereafter, suggesting the authors were not able to verify their results.

For a decade, the telescopes of the Arizona Radio Observatory (ARO) have conducted a confusion-limited spectral-line survey of Sgr B2(N), the hot core of a giant molecular cloud near the Galactic center (e.g. Nummelin et al. 1998). The survey encompasses the frequency ranges 68–116, 130–172, and 210–280 GHz in the 3, 2, and 1 mm atmospheric windows (D. T. Halfen et al. in preparation), with sensitivities of 3–5 mK (1σ). Sgr B2(N) contains at least 70 different chemical species, such as methyl amine and acetamide (e.g. Halfen, Ilyushin, & Ziurys 2011, 2013). The ARO survey provides a unique dataset for identifying larger organic-type molecules, particularly asymmetric top species that are known for their spectral complexity. The survey enables a complete view of the mm-wave spectrum of a given asymmetric top, which, because of the high density of lines in Sgr B2(N), are subject to false identifications (e.g. Snyder et al. 2005; Apponi et al. 2006). The high probability of chance coincidences with transitions from other molecules in this source makes new, definitive identifications difficult, but the global picture of a survey establishes a much higher confidence level. Furthermore, the velocity structure of a given source like Sgr B2(N) can be comprehensively analyzed. The ARO survey data has shown that ~70% of the over 70 species thus far identified in the 1–3 mm range have two distinct velocity components at 64 ± 2, and 73 ± 2 km $s^{-1}$, respectively, each with a characteristic line widths of 10 ± 2 km $s^{-1}$ (Halfen et al., in prep.).

In this letter, we present the first definitive detection of a new interstellar molecule, $CH_3NCO$, towards Sgr B2(N), using the 3 mm data of the ARO survey. $CH_3NCO$ is the next step in complexity following HNCO and HOCN, both also identified in this source (e.g. Brünken et al. 2010). The search was based on a new spectroscopic analysis of the previous data and new Fourier Transform Millimeter-Wave (FTmmW) measurements, using a rho-axis method Hamiltonian with the RAM36 program (Ilyushin et al. 2010). All favorable asymmetry and internal rotation components of $CH_3NCO$ from five consecutive rotational transitions from 68–105 GHz were detected, with 57% of the lines being uncontaminated. The spectral lines show the typical velocity structure of Sgr B2(N). Here we present these results.

## 2. Spectroscopy of Methyl Isocyanate

The rotational spectrum of $CH_3NCO$ has long been a challenge for experimentalists. The NCO group lies off the molecular symmetry axis at an angle of ~140°, creating an asymmetric top species with a low-lying (~172 $cm^{-1}$) C-N-C bending mode (Lett & Flygare 1967; Koput 1984). In addition, the barrier to the methyl group internal rotation is estimated to be quite low (~20 $cm^{-1}$), such that the $CH_3$ moiety almost freely spins about the symmetry axis. The pure rotational spectrum is therefore quite complex with A and E torsional states and the presence of strong vibrational satellite lines. Early laboratory spectroscopic measurements of $CH_3NCO$ were performed by Curl et al. (1963) and Lett & Flygare (1967), who recorded microwave data for the low *J* *a*-type transitions below 35 GHz. Asymmetry doubling, labeled with quantum number $K_a$, and internal rotation interactions, indicated by quantum number *m*, as well as nitrogen quadrupole hyperfine structure, all contributed to the complex spectra recorded by these authors. Neither group was able to successfully model their data with a rotational Hamiltonian. Additional work was done by Kasten & Dreizler (1986) with Fourier transform microwave (FTMW) spectroscopy from 8–17 GHz. Here only the quadrupole hyperfine splittings were analyzed.

A major step forward in unraveling the spectrum of $CH_3NCO$ was made by Koput in 1986. In this work, the rotational spectrum was measured from 8–40 GHz with microwave techniques. More than 200 lines were analyzed using the five-dimensional quasi-symmetric top Hamiltonian from the 1984 work by this author, with a resulting *rms* of 3 MHz including the torsional and C-N-C bending mode data. Soon thereafter, Koput (1988) determined the *a*-dipole moment of $CH_3NCO$ to be $\mu_a$ = 2.885 D, in close agreement with Kasten and Dreizler ($\mu_a$ = 2.882 D). More recently, Palmer & Nelson (2003) calculated the structure of $CH_3NCO$ with B3LYP, MP2, and CCSD(T) methods, finding it consistent with the microwave studies.

In order to establish accurate rest frequencies in the 3 mm window, which lies above 40

GHz, we conducted a new analysis of $CH_3NCO$ using the previous data of Koput (1986) and new spectra measured in the Ziurys laboratory in the 60–88 GHz range. The rho-axis method Hamiltonian with the RAM36 program was employed for this analysis (Ilyushin et al. 2010, 2013). This Hamiltonian is particularly useful because it incorporates torsional motion for a nearly free methyl rotor, as found for $CH_3NCO$. The new laboratory work was carried out in the frequency range 60–88 GHz with the FTmmW spectrometer; $CH_3NCO$, diluted in argon, was introduced into the system in a supersonic jet expansion (see Sun et al. 2009 for details). The $J = 1 \leftarrow 0$ through $J = 4 \leftarrow 3$ transitions up to $K_a = 3$ in $v_b = 0$ (CNC bending mode) from all the $m = 0$ and ±3 (A state) and all $m = 1$, -2, and 4 transitions (E state) from Koput (1986) were fit, along with twenty-one lines from the $J = 7 \rightarrow 6$ to $10 \rightarrow 9$ transitions with $K_a = 0$ and 1 for the A ($m = 0$) and E ($m = 1$) states; see Table 1. The subset of data was selected from Koput with the aim of minimizing the possible influence of torsion-vibration interactions with the low-lying CNC bending mode.

Spectroscopic constants were determined from an analysis of 96 total lines and given in Table 2. For an explanation of the parameters, see Ilyushin et al. (2010). The main parameters of the RAM Hamiltonian agree well with the values recalculated from Koput (1986), as shown in Table 2.

From these constants, rest frequencies for $CH_3NCO$ were predicted from 68–105 GHz for the $K_a = 0$ and 1 asymmetry components in the $J = 8 \rightarrow 7$ through $J = 12 \rightarrow 11$ transitions in both A and E states in $v_b = 0$. These lines all lie <60 K above the ground state. The five consecutive transitions in both the E and A states in the harmonic $K_a = 0$ components, as well as the $K_a = 1$ doublets, generate a unique fingerprint pattern that is far more distinctive than random $K_a$ or $K_c$ components of varying energies. For the lines that had predicted frequencies, uncertainties are ≤1 MHz, which is more than sufficient for identification in Sgr B2(N). The fourteen predicted and the sixteen measured transition frequencies, their uncertainties, upper state energies ($E_u$), and line strengths ($\mu^2$S) in the 68–105 GHz range are listed in Table 3.

## 3. Observations

Observations of $CH_3NCO$ were performed as part of the spectral survey of Sgr B2(N). The data were collected during Sept. 2002 – Mar. 2014 using the ARO 12 m telescope on Kitt Peak, Arizona and the Submillimeter Telescope (SMT) on Mount Graham. For the 3 mm observations (68–116 GHz), the receivers employed dual-polarization, single-sideband (SSB) SIS mixers. Initially, SIS mixers were used with typical image rejection of ≥16 dB, obtained by tuning the backshorts. Then observations were taken using a newer dual-polarization receiver containing ALMA Band 3 (83–116 GHz) sideband-separating (SBS) mixers, where the usual image rejection was ~20 dB, attained within the mixer. The backend used was a 600 MHz bandwidth millimeter autocorrelator (MAC) with either 390 kHz or 781 kHz resolution; all data were then smoothed to 1 MHz resolution. The backend was configured in parallel mode to accommodate both receiver channels.

The temperature scale was established using the chopper wheel method, corrected for forward spillover losses, and given as $T_R^*$. The radiation temperature $T_R$ is then $T_R = T_R^*/\eta_c$, where $\eta_c$ is the corrected beam efficiency. All measurements were obtained in position-switching mode toward Sgr B2(N) ($\alpha = 17^h44^m09.5^s$; $\delta = -28°21'20''$; B1950) with an OFF position 30′ west in azimuth. Image contamination was determined by direct observation of the image sideband and a 10 MHz shift of the local oscillator. The telescope pointing was established using observations of planets and quasars. Telescope parameters are given in Table 3.

## 4. Results

Emission was observed at all the 30 favorable transitions of $CH_3NCO$ in the 3 mm band of the Sgr B2(N) survey. Seventeen of these transitions had clean, resolvable spectral lines, as shown in Figure 1, while the remaining thirteen were contaminated by other molecules. These identifications are not at random frequencies. The $K_a = 0$ component of the A state was clearly detected in all five successive $J + 1 \rightarrow J$ rotational transitions (see Figure 1a). This series of lines is harmonic in frequency, and their appearance bolsters the case for the identification. For the E state, two $K_a = 0$ lines were identified, as the other three were contaminated; see Table 3 and Figure 1. Also identified were most of the $K_a = 1$ asymmetry components for the A state. These lines appear as widely-spaced (~1.5 GHz) doublets for a given transition $J + 1 \rightarrow J$, centered with respect to the $K_a = 0$ line and are another important aspect of the "fingerprint." For the $J = 9 \rightarrow 8$, $11 \rightarrow 10$, and $12 \rightarrow 11$ A transitions in $CH_3NCO$, this asymmetry doublet has clearly been detected (see Figure 1a,b); for the two other transitions, one of the doublets is obscured by other species (see Table 3). For the E state, two uncontaminated $K_a = \pm1$ lines ($J = 9 \rightarrow 8$ and $J = 12 \rightarrow 11$) were observed (Figure 1b). The spectral features have consistent intensities in the range $T_R^* \sim 20$–70 mK.

For the $J = 9 \rightarrow 8$ transition, the $K_a = 0$ line and $K_a = 1$ doublets of the A and E states were all identified, with the exception of the $9_{-1,9} \rightarrow 8_{-1,8}$ E line, which is obscured by $HC_5N$. These five transitions illustrate the distinctive spectral pattern of $CH_3NCO$, as shown in the inset in Figure 2. It is highly unlikely then that the lines attributed to $CH_3NCO$ are a result of chance coincidences. It should also be noted that none of the observed $CH_3NCO$ frequencies can be accounted for by favorable transitions of other known molecules in Sgr B2(N).

The $CH_3NCO$ line profiles also exhibit the typical two velocity components observed for most molecules in Sgr B2(N), including HNCO (e.g. Belloche et al. 2013; Halfen et al., in prep.). Although the lines are weak, the $CH_3NCO$ features have a "main" velocity component near $V_{LSR}$ ≈ 62 km $s^{-1}$ with a "shoulder" near 73 km $s^{-1}$. Both features can be modeled with a linewidth of $\Delta V_{1/2} \approx$ 10 km $s^{-1}$. At the lower end of the 3 mm band ≤70 GHz, the spectral resolution is lower in terms of velocity such that the two features merge into a single line with roughly twice the linewidth (~20 km $s^{-1}$), such as the $J_{K_a,K_c} = 8_{0,8} \rightarrow 7_{0,7}$ A transition. The two velocity components are apparent in most transitions, such as the $J_{K_a,K_c} = 9_{0,9} \rightarrow 8_{0,8}$ A, $9_{1,9} \rightarrow 8_{1,8}$ A, $11_{0,11} \rightarrow 10_{0,10}$ A, $12_{1,12} \rightarrow 11_{1,11}$ A and $12_{0,12} \rightarrow 11_{0,11}$ A lines. Note that the $J_{K_a,K_c} = 10_{0,10} \rightarrow 9_{0,9}$ A and E lines are partially blended together. Also, the 73 km $s^{-1}$ velocity component of the $10_{1,9} \rightarrow 9_{1,8}$ A transition is partially confused with a U line, but the 62 km $s^{-1}$ feature is clean. In the $11_{1,11} \rightarrow 10_{1,10}$ A line, the 62 km $s^{-1}$ feature is partially blended, while that at 73 km $s^{-1}$ is uncontaminated.

Gaussian profiles were fit to the observed features and the resulting line parameters are given in Table 3. The overall intensities of the uncontaminated lines range from $T_R^* = 22–73$ mK for the 62 km $s^{-1}$ component and $T_R^* = 21–56$ mK for the 73 km $s^{-1}$ lines. The line widths for each velocity component were modeled as 10 km $s^{-1}$, and produce a feature with an overall width of about 19–22 km $s^{-1}$.

## 5. Discussion

### *5.1 Column Densities and Fractional Abundances of $CH_3NCO$*

Column densities for $CH_3NCO$ were derived using two techniques. First, the rotational diagram method was used (e.g. Turner 1991). This analysis was carried out assuming both source components fill the main telescope beam. For the lowest frequency transitions where the two velocity components are not well resolved, the intensity used was split in a 2:1 ratio for the 62 and 73 km s$^{-1}$ components, respectively. A very similar intensity ratio is observed in multiple species towards Sgr B2(N) (Halfen et al., in prep.). The resulting diagram is shown in Figure 2. The points on the diagram are extremely consistent within the errors. From the analysis, the rotational temperature and column density of $CH_3NCO$ were determined to be $T_{\rm rot} = 24 \pm 3$ K and $N_{\rm tot} = 2.3 \pm 0.4 \times 10^{13}$ cm$^{-2}$ for the 62 km s$^{-1}$ component; for the features at 73 km s$^{-1}$, $T_{\rm rot} = 28 \pm 5$ K and $N_{\rm tot} = 1.5 \pm 0.3 \times 10^{13}$ cm$^{-2}$ was derived. Assuming a column density for $H_2$ of $N_{\rm tot} = 3 \times 10^{24}$ cm$^{-2}$ for Sgr B2(N) (Nummelin et al. 2000), the fractional abundances of $CH_3NCO$ are $f(X/H_2) = 7.6 \times 10^{-12}$ and $5.0 \times 10^{-12}$ for the 62 and 73 km s$^{-1}$ components, respectively.

The second analysis method employed an LTE code to model the observed spectra of $CH_3NCO$. The code assumes all lines are optically thin and the source fills the telescope beams. The line profiles were modeled with line widths of 10 km s$^{-1}$ and two LSR components at 62 and 73 km s$^{-1}$. The best fit to the observed data was achieved assuming $T_K = 25$ K and column densities of $2.3 \times 10^{13}$ cm$^{-2}$ and $1.5 \times 10^{13}$ cm$^{-2}$ for the 62 and 73 km s$^{-1}$ components, respectively. The predicted line profiles are superimposed in red over the actual data in Figure 1. There is very good agreement between the model and observed spectra. Also, the values of the column densities from both methods are in good agreement, and $T_K \approx T_{rot}$, as expected at LTE. From the derived column densities, fractional abundances are $f(X/H_2) = 7.6 \times 10^{-12}$ and $5.0 \times 10^{-12}$ for the 62 and 73 km s$^{-1}$ components, respectively. These abundances are comparable to other organic molecules, such as $CH_3CONH_2$ and $CHOCH_2OH$ (e.g. Halfen et al. 2011).

*5.2 Comparison with HNCO and HOCN*

From the ARO Sgr B2(N) survey spectra, the column density and rotational temperature of HNCO has been determined from LTE modeling. There appears to be low temperature gas with two velocity components at $V_{LSR} \approx 62$ and 73 km s$^{-1}$, both with $\Delta V_{1/2} \approx 12$ km s$^{-1}$, $T_{rot} \approx 16$ K, both with $N_{tot} \approx 8.0 \times 10^{14}$ cm$^{-2}$. The line parameters and rotational temperature of the low temperature component of HNCO are very similar to those of $CH_3NCO$, suggesting they are related chemically. (Note that higher temperature material with $T_{rot} \approx 123$ K is also apparent in the HNCO data.) The fractional abundances for the lower temperature HNCO are $f(X/H_2) \approx 2.7 \times 10^{-10}$ in each velocity component, resulting in an abundance ratio of $CH_3NCO$/HNCO is ~1/35 and ~1/53, for 62 and 73 km s$^{-1}$ features, respectively.

HOCN, the higher-energy isomer of HNCO, also has the 62 km s$^{-1}$ and 73 km s$^{-1}$ velocity components found in $CH_3NCO$, with $\Delta V_{1/2} \approx 12$ km s$^{-1}$. For this species, the LTE model yielded $T_{rot} \approx 10$ K and $N_{tot} \approx 5.0 \times 10^{12}$ cm$^{-2}$ for the 62 km s$^{-1}$ component, and $T_{rot} \approx 8.0$ K, and $N_{tot} \approx 5.0 \times 10^{12}$ cm$^{-2}$ for the second component. Fractional abundances are $f(X/H_2) \approx 1.7 \times 10^{-12}$ for both velocity components (Halfen et al., in prep.), implying $CH_3NCO$/HOCN ≈ 4/1.

While the former molecule is more abundant, both HNCO and HOCN could produce $CH_3NCO$ in the gas-phase via reactions with $CH_3$:

$$HNCO + CH_3 \rightarrow CH_3NCO + H \quad (1)$$

$$HOCN + CH_3 \rightarrow CH_3NCO + H \quad (2)$$

Because $CH_3$ is a radical, these processes could have rates as high as ~$10^{-10}$ cm$^3$ s$^{-1}$ (Woodall et al. 2007); based on the heats of formation (Quan et al. 2010), however, Equation (1) is endothermic, while Equation (2) is exothermic. Another reasonable formation route for $CH_3NCO$ is the reaction of HNCO or HOCN with protonated methane ($CH_5^+$), followed by dissociative electron recombination:

$$HNCO + CH_5^+ \rightarrow CH_3NCOH^+ + H_2 \quad (3)$$

$$HOCN + CH_5^+ \rightarrow CH_3NCOH^+ + H_2 \quad (3)$$

$$CH_3NCOH^+ + e^- \rightarrow CH_3NCO + H \quad (4)$$

The initial ion–molecule reactions have never been studied experimentally, but could proceed near the Langevin rate of ~$10^{-9}$ $cm^3$ $s^{-1}$ (Landau & Lifshitz 1965).

*5.3 In Relation to Comets*

In comet 67P, the COSAC mass spectrometer detected four new organic cometary compounds in the gas phase, which were sublimated from excavated material from the first false landing, including $CH_3NCO$. This species was found at the level of 1.3% relative to water (Goesmann et al. 2015). Of the 16 molecules identified in the COSAC spectrum, $CH_3NCO$ was the third most abundant following $H_2O$ (100%) and formamide, $NH_2CHO$ (1.8%). Interestingly, HNCO was found to have a 0.3% abundance relative to $H_2O$ in comet 67P, suggesting a $CH_3NCO$/HNCO ratio of ~4.

In Sgr B2(N), $H_2O$ is estimated to have a column density of $9 \pm 3 \times 10^{16}$ $cm^{-2}$ (Cernicharo et al. 2006), implying that $CH_3NCO$ is ~0.04% of the water abundance, summing over both velocity components. This percentage is about a factor of 30 less than that in comet 67P. In contrast, the $NH_2CHO/CH_3NCO$ ratio in Sgr B2(N) is ~5, based on the column density of the "cold" component from Halfen et al. (2011). This value is remarkably close to the ratio of ~1.4 found in the comet.

This research was supported by NSF grants AST-1140030 (University Radio Observatories), AST-1211502 and by the NASA NExSS program via grant NNX13ZDA017C.

**References**

Apponi, A. J., Halfen, D. T., Ziurys, L. M., Hollis, J. M., Remijan, A. J., & Lovas, F. J. 2006, ApJ, **643**, L29

Belloche, A., Müller, H. S. P., Menten, K. M., Schilke, P., & Comito, C. 2013, A&A, **559**, A47

Biver, N., et al. 2014, A&A, **566**, L5

Brünken, S., Belloche, A., Martín, S., Verheyen, L., & Menten, K. M. 2010, A&A, **516**, A109

Cernicharo, J., Goicoechea, J. R., Pardo, J. R., & Asensio-Ramos, A. 2006, ApJ, **642**, 940

Crovisier, J., Bockelée-Morvan, D., Colom, P., et al. 2004, A&A, **418**, 1141

Curl, R. F., Jr., Rao, V. M., Sastry, K. V. L. N., & Hodgeson, J. A. 1963, J. Chem. Phys., **39**, 3335

Goesmann, F., et al. 2015, Science, **349**, aab0689

Halfen, D. T., Ilyushin, V. V., & Ziurys, L. M. 2011, ApJ, **743**, 60

Halfen, D. T., Ilyushin, V. V., & Ziurys, L. M. 2013 ApJ, **767**, 66

Hartogh, P., Lis, D.C., Bockelée-Morvan, D., et al. 2011, Nature, **478**, 218

Ilyushin, V. V., Kisiel, Z., Pszczólkowski, L., Mäder, H., & Hougen, J. T. 2010, J. Mol. Spectrosc., **259**, 26

Ilyushin V. V.,. Endres C. P, Lewen F., Schlemmer S., Drouin B. J. 2013, J. Mol. Spectrosc., **290**, 31

Kasten, W., & Dreizler, H. 1986, Z. Natur., **41a**, 637

Koput, J. 1984, J. Mol. Spectrosc., **106**, 12

Koput, J. 1986, J. Mol. Spectrosc., **115**, 131

Koput, J. 1988, J. Mol. Spectrosc., **127**, 51

Landau, L. D., & Lifshitz, E. M. 1965, Quantum Mechanics (Oxford: Pergamon)

Lett, R. G., & Flygare, W. H. 1967, J. Chem. Phys., **47**, 4730

Mumma, M. J., & Charnley, S. B. 2011, ARA&A, **49**, 471

Nummelin, A. et al. 1998, ApJS, **117**, 427

Nummelin, A. et al. 2000, ApJS, **128**, 213

Palmer, M. H., & Nelson, A. D. 2003, J. Mol. Struct., **660**, 49

Quan, D., Herbst, E., Osamura, Y., & Roueff, E. 2010, ApJ, **725**, 2101

Pulliam, R. L., Remijan, A. J., & Loomis, R. A. 2013, AAS Meeting #221, #352.11

Sun, M., Apponi, A. J., & Ziurys, L. M. 2009, J. Chem. Phys., **130**, 034309

Snyder, L. E., Lovas, F. J., Hollis, J. M., Friedel, D. N., Jewell, P. R., Remijan, A., Ilyushin, V. V., Alekseev, E. A., & Dyubko, S. F. 2005, ApJ, **619**, 914

Turner, B. E. 1991, ApJS, **76**, 617

Woodall, J., Agúndez, M., Markwick-Kemper, A. J., & Millar, T. J. 2007, A&A, **466**, 1197

Table 1
Previous and New Rest Frequencies of $CH_3NCO$ ($v_b$ = 0) of the Revised Analysis

| $J'$ | $K_a'$ | $K_c'$ | ↔ | $J''$ | $K_a''$ | $K_c''$ | Sym | $m$ | $\nu_{obs}$ | $\nu_{obs}$ - $\nu_{calc}$ |
|---|---|---|---|---|---|---|---|---|---|---|
| 1 | 0 | 1 | ← | 0 | 0 | 0 | A | 0 | 8671.369 [a] | 0.234 |
| 1 | 0 | 1 | ← | 0 | 0 | 0 | E | -2 | 8712.440 | -0.006 |
| 1 | 0 | 1 | ← | 0 | 0 | 0 | E | 1 | 8714.600 | 0.248 |
| 1 | 0 | 1 | ← | 0 | 0 | 0 | A | 3 | 8722.070 | -0.037 |
| 1 | 0 | 1 | ← | 0 | 0 | 0 | A | -3 | 8722.930 | 0.289 |
| 1 | 0 | 1 | ← | 0 | 0 | 0 | E | 4 | 8737.510 | 0.172 |
| | | | | | | | | | | |
| 9 | 1 | 8 | → | 8 | 1 | 7 | E | 1 | 77107.478 | 0.010 |
| 9 | -1 | 9 | → | 8 | -1 | 8 | E | 1 | 77211.707 | -0.357 |
| 9 | 1 | 8 | → | 8 | 1 | 7 | A | 0 | 77347.355 | 0.438 |
| 9 | 0 | 9 | → | 8 | 0 | 8 | A | 0 | 78017.932 | 0.309 |
| 9 | 0 | 9 | → | 8 | 0 | 8 | E | 1 | 78087.673 | 0.456 |
| 9 | 1 | 9 | → | 8 | 1 | 8 | A | 0 | 78758.868 | 0.357 |

**Notes.** Frequencies in MHz.
Below 40 GHz, frequencies from Koput (1986), above from this work, unless specified.
[a] From Kasten & Dreizler (1986).
[b] Estimated from observed spectra in Figure 1.

(This table is available in its entirety in a machine-readable form in the online journal. A portion is shown here for guidance regarding its form and content.)

Table 2
Revised Spectroscopic Constants for $CH_3NCO$ ($\tilde{X}^1A'$)

| Parameter | This work | Previous work [a] |
|---|---|---|
| $V_3$ ($cm^{-1}$) | 21.76(14) | 20.72(30) |
| $F$ ($cm^{-1}$) | 8.769(40) | 8.58 |
| $\rho$ (unitless) | 0.44814(37) | 0.406 |
| $A_{RAM}$ | 78395(410) | 72054 |
| $B_{RAM}$ | 4442.982(49) | 4432.7 |
| $C_{RAM}$ | 4256.691(31) | 4240.8 |
| $D_{ab}$ | -1918(14) | -1906 |
| $D_J$ | 0.001676(69) | |
| $D_{JK}$ | -1.339(12) | |
| $d_J$ | 0.00062(10) | |
| $F_J$ | 1.283(18) | |
| $F_{JK}$ | -0.0401(74) | |
| $F_{mJ}$ | -0.0075(11) | |
| $\rho_{mJ}$ | -0.0267(34) | |
| $\rho_J$ | 2.691(36) | |
| $\rho_m$ | -171(34) | |
| $V_{3ab}$ | 112(33) | |
| $V_{3K}$ | -3633(510) | |
| *rms* | 0.653 | |

**Notes.** Parameters are in MHz, unless otherwise specified.
Errors quoted are 1σ in the last quoted digits.
[a] Recalculated from Koput (1986).

Table 3
Observed Parameters for $CH_3NCO$ in Sgr B2(N)

| Transition $J'_{K_a,K_c} \rightarrow J''_{K_a,K_c}$ | Sym | $m$ | Frequency (MHz) | $E_u$ (K) | $\mu^2S$ ($D^2$) | $\eta_c$ | $\theta_b$ (″) | $T_R^*$ (K) | $\Delta V_{1/2}$ (km/s) | $V_{LSR}$ (km/s) | Comment |
|---|---|---|---|---|---|---|---|---|---|---|---|
| $8_{1,7} \rightarrow 7_{1,6}$ | E | 1 | 68536.76(0.34) | 40.14 | 63.00 | 0.94 | 92 | ~0.04 | ~22 | ~64 | Blended with U line |
| $8_{-1,8} \rightarrow 7_{-1,7}$ | E | 1 | 68617.550(0.005) | 22.06 | 63.64 | 0.94 | 92 | ~0.06 | ~22 | ~64 | Blended with U line |
| $8_{1,7} \rightarrow 7_{1,6}$ | A | 0 | 68753.890(0.005) | 20.75 | 64.59 | 0.94 | 91 | --- | --- | --- | Blended with $CH_3NH_2$ |
| $8_{0,8} \rightarrow 7_{0,7}$ | A | 0 | 69353.888(0.020) | 14.95 | 65.61 | 0.94 | 91 | 0.057 ± 0.004 | 22 ± 4 | 66 ± 4 | Clean line, blended velocity components |
| $8_{0,8} \rightarrow 7_{0,7}$ | E | 1 | 69469.196(0.005) | 27.18 | 66.08 | 0.94 | 90 | ~0.04 | ~20 | ~63 | Shoulder on $SO_2$ line |
| $8_{1,8} \rightarrow 7_{1,7}$ | A | 0 | 70009.179(0.005) | 21.03 | 64.59 | 0.94 | 90 | 0.032 ± 0.004 | 21 ± 4 | 63 ± 4 | Clean line, blended velocity components |
| $9_{1,8} \rightarrow 8_{1,7}$ | E | 1 | 77107.478(0.005) | 43.85 | 71.11 | 0.93 | 82 | 0.022 ± 0.002 | 10 ± 4 | 62 ± 4 | Clean line |
| | | | | | | | | --- | --- | 73 [a] | Blended with c-$C_3H_2$ |
| $9_{-1,9} \rightarrow 8_{-1,8}$ | E | 1 | 77211.707(0.005) | 25.76 | 71.85 | 0.93 | 81 | --- | --- | --- | Blended with $HC_5N$ |
| $9_{1,8} \rightarrow 8_{1,7}$ | A | 0 | 77347.355(0.005) | 24.47 | 72.90 | 0.93 | 81 | 0.036 ± 0.004 | 10 ± 4 | 62 ± 4 | Clean line, two velocity |
| | | | | | | | | 0.029 ± 0.004 | 10 ± 4 | 73 ± 4 | components |
| $9_{0,9} \rightarrow 8_{0,8}$ | A | 0 | 78017.932(0.020) | 18.69 | 73.82 | 0.93 | 81 | 0.070 ± 0.005 | 10 ± 2 | 62 ± 2 | Clean line, two velocity |
| | | | | | | | | 0.056 ± 0.005 | 10 ± 2 | 73 ± 2 | components |
| $9_{0,9} \rightarrow 8_{0,8}$ | E | 1 | 78087.673(0.005) | 30.93 | 74.24 | 0.93 | 81 | 0.030 ± 0.005 | 10 ± 2 | 62 ± 2 | Clean line, two velocity |
| | | | | | | | | 0.021 ± 0.005 | 10 ± 2 | 73 ± 2 | components |
| $9_{1,9} \rightarrow 8_{1,8}$ | A | 0 | 78758.868(0.005) | 24.81 | 72.90 | 0.93 | 80 | 0.037 ± 0.004 | 10 ± 3 | 62 ± 3 | Clean line, two velocity |
| | | | | | | | | 0.025 ± 0.004 | 10 ± 3 | 73 ± 3 | components |
| $10_{1,9} \rightarrow 9_{1,8}$ | E | 1 | 85679.63(0.44) | 47.96 | 79.20 | 0.91 | 73 | --- | --- | --- | Blended with H42α |
| $10_{-1,10} \rightarrow 9_{-1,9}$ | E | 1 | 85811.557(0.005) | 29.89 | 80.04 | 0.91 | 73 | --- | --- | --- | Blended with $CH_2CHCN$ $v_{15} = 1$ |
| $10_{1,9} \rightarrow 9_{1,8}$ | A | 0 | 85938.967(0.005) | 28.60 | 81.19 | 0.91 | 73 | 0.044 ± 0.004 | 10 ± 3 | 62 ± 3 | Clean line |
| | | | | | | | | --- | --- | 73 [a] | Blended with U line |
| $10_{0,10} \rightarrow 9_{0,9}$ | A | 0 | 86680.195(0.005) | 22.86 | 82.02 | 0.91 | 73 | 0.062 ± 0.003 | 10 ± 3 | 62 ± 3 | Clean line, two velocity |
| | | | | | | | | 0.044 ± 0.003 | 10 ± 3 | 73 ± 3 | components |
| $10_{0,10} \rightarrow 9_{0,9}$ | E | 1 | 86686.575(0.005) | 35.09 | 82.38 | 0.91 | 73 | 0.045 ± 0.003 | 10 ± 3 | 62 ± 3 | Clean line |
| | | | | | | | | ~0.03 | ~10 | 73 [a] | Partially blended with A state |
| $10_{1,10} \rightarrow 9_{1,9}$ | A | 0 | 87506.605(0.005) | 29.01 | 81.19 | 0.91 | 72 | --- | --- | --- | Blended with $CH_3NH_2$ |
| $11_{1,10} \rightarrow 10_{1,9}$ | E | 1 | 94253.49(0.74) | 52.49 | 87.28 | 0.89 | 67 | --- | --- | --- | Blended with $NH_2CHO$ |
| $11_{-1,11} \rightarrow 10_{-1,10}$ | E | 1 | 94415.21(0.45) | 34.42 | 88.23 | 0.89 | 67 | --- | --- | --- | Blended with $^{13}CH_3OH$ |
| $11_{1,10} \rightarrow 10_{1,9}$ | A | 0 | 94530.69(0.42) | 33.14 | 89.47 | 0.89 | 67 | 0.050 ± 0.004 | 10 ± 2 | 62 ± 2 | Clean line, two velocity |
| | | | | | | | | 0.026 ± 0.004 | 10 ± 2 | 73 ± 2 | components |
| $11_{0,11} \rightarrow 10_{0,10}$ | E | 1 | 95268.79(0.40) | 39.67 | 90.50 | 0.89 | 66 | --- | --- | --- | Blended with $^{13}CH_3OH$ |
| $11_{0,11} \rightarrow 10_{0,10}$ | A | 0 | 95341.15(0.36) | 27.44 | 90.22 | 0.89 | 66 | 0.065 ± 0.005 | 10 ± 3 | 62 ± 3 | Clean line, two velocity |
| | | | | | | | | 0.043 ± 0.005 | 10 ± 3 | 73 ± 3 | components |
| $11_{1,11} \rightarrow 10_{1,10}$ | A | 0 | 96253.95(0.48) | 33.63 | 89.47 | 0.89 | 65 | ~0.05 | ~10 | 62 [a] | Blended with $(CH_2OH)_2$ |
| | | | | | | | | 0.029 ± 0.003 | 10 ± 2 | 73 ± 2 | Clean line |
| $12_{1,11} \rightarrow 11_{1,10}$ | E | 1 | 102829.29(1.15) | 57.43 | 95.34 | 0.87 | 61 | ~0.01 | --- | --- | Blended with U lines |
| $12_{-1,12} \rightarrow 11_{-1,11}$ | E | 1 | 103023.61(0.71) | 39.37 | 96.41 | 0.87 | 61 | 0.047 ± 0.002 | 10 ± 3 | 62 ± 3 | Clean line, two velocity |
| | | | | | | | | 0.037 ± 0.002 | 10 ± 3 | 73 ± 3 | components |
| $12_{1,11} \rightarrow 11_{1,10}$ | A | 0 | 103121.61(0.67) | 38.09 | 97.74 | 0.87 | 61 | 0.048 ± 0.003 | 10 ± 3 | 62 ± 3 | Clean line, two velocity |
| | | | | | | | | 0.034 ± 0.003 | 10 ± 2 | 73 ± 2 | components |
| $12_{0,12} \rightarrow 11_{0,11}$ | E | 1 | 103835.02(0.59) | 44.66 | 98.60 | 0.87 | 61 | --- | --- | --- | Blended with $CH_2CHCN$ $v_{11} = 1$, $HC_5N$ |
| $12_{0,12} \rightarrow 11_{0,11}$ | A | 0 | 103999.94(0.50) | 32.43 | 98.42 | 0.87 | 60 | 0.073 ± 0.004 | 10 ± 2 | 62 ± 2 | Clean line, two velocity |
| | | | | | | | | 0.044 ± 0.004 | 10 ± 2 | 73 ± 2 | components |
| $12_{1,12} \rightarrow 11_{1,11}$ | A | 0 | 105000.26(0.79) | 38.68 | 97.73 | 0.87 | 60 | 0.041 ± 0.004 | 10 ± 2 | 62 ± 2 | Clean line, two velocity |
| | | | | | | | | 0.025 ± 0.004 | 10 ± 2 | 73 ± 2 | components |

**Notes.** Source coordinates (B1950): $\alpha = 17^h44^m09.^s5$, $\delta = -28^o21'20''$. Spectra were observed with the Millimeter Auto Correlator (MAC), and re-sampled at 1 MHz resolution using a standard cubic spline algorithm. Line parameters were derived from a least-squares fit of a Gaussian profile to the spectrum. Measured frequencies have errors (3σ) of 5–20 kHz, while those for the predicted frequencies are ≤1 MHz.

[a] Assumed value

**Figure Captions**

Figure 1. – Detected lines of $CH_3NCO$ measured towards Sgr B2(N) with the ARO 12 m telescope at 3 mm. Rest frequencies and transitions are indicated on the spectra, assuming an LSR velocity of 64 km $s^{-1}$. The spectral resolution is 1 MHz. The modeled spectrum overlays the data in red, assuming two well-justified velocity components (62 and 73 km $s^{-1}$) and $T_{rot}$ = 25 K; see text. Figure 1a displays the five successive uncontaminated $K_a$ = 0 transitions of the A species, as well as the two $K_a$ = 0 E state lines, one which appears as shoulder on the $10_{0,10} \rightarrow 9_{0,9}$ A profile, as indicated by arrows. The lower two panels of Figure 1a and all of Figure 1b shows the distinguishable $K_a$ = 1 lines from both A and E states. There is some partial blending with $(CH_2OH)_2$ and a U line for the $11_{1,11} \rightarrow 10_{1,10}$ and $10_{1,9} \rightarrow 9_{1,8}$ A transitions, and the $12_{1,11} \rightarrow 11_{1,10}$ A line lies adjacent to a $CH_2CH^{13}CN$ feature. Approximate integration time per spectrum is 4–16 hours.

Figure 2. – Rotational temperature diagrams of the 62 and 73 km $s^{-1}$ velocity components of $CH_3NCO$ in Sgr B2(N) with $3\sigma$ error bars, along with resulting column densities and rotational temperatures. The inset displays the observed vs. predicted $K_a$ component pattern of the $J = 9 \rightarrow 8$ transition - the distinguishing "fingerprint" for $CH_3NCO$.

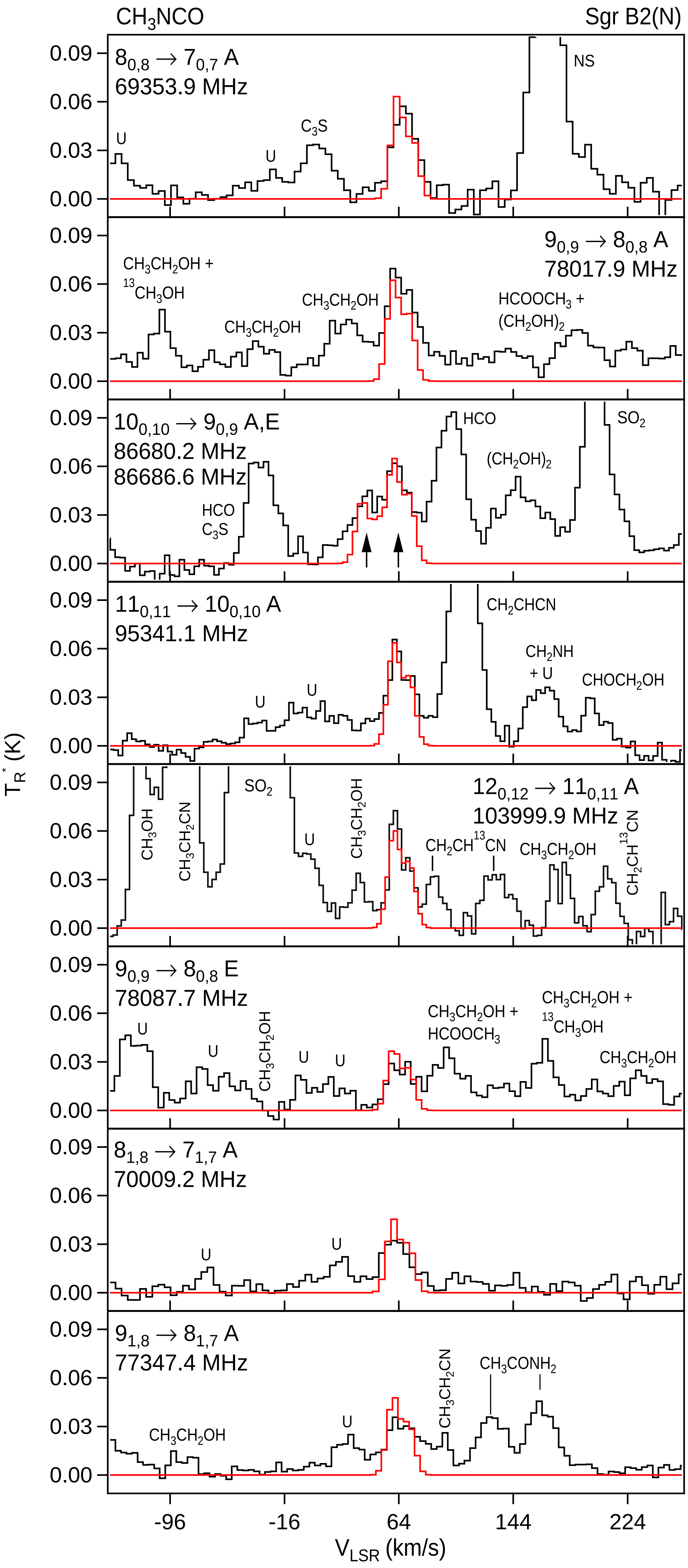

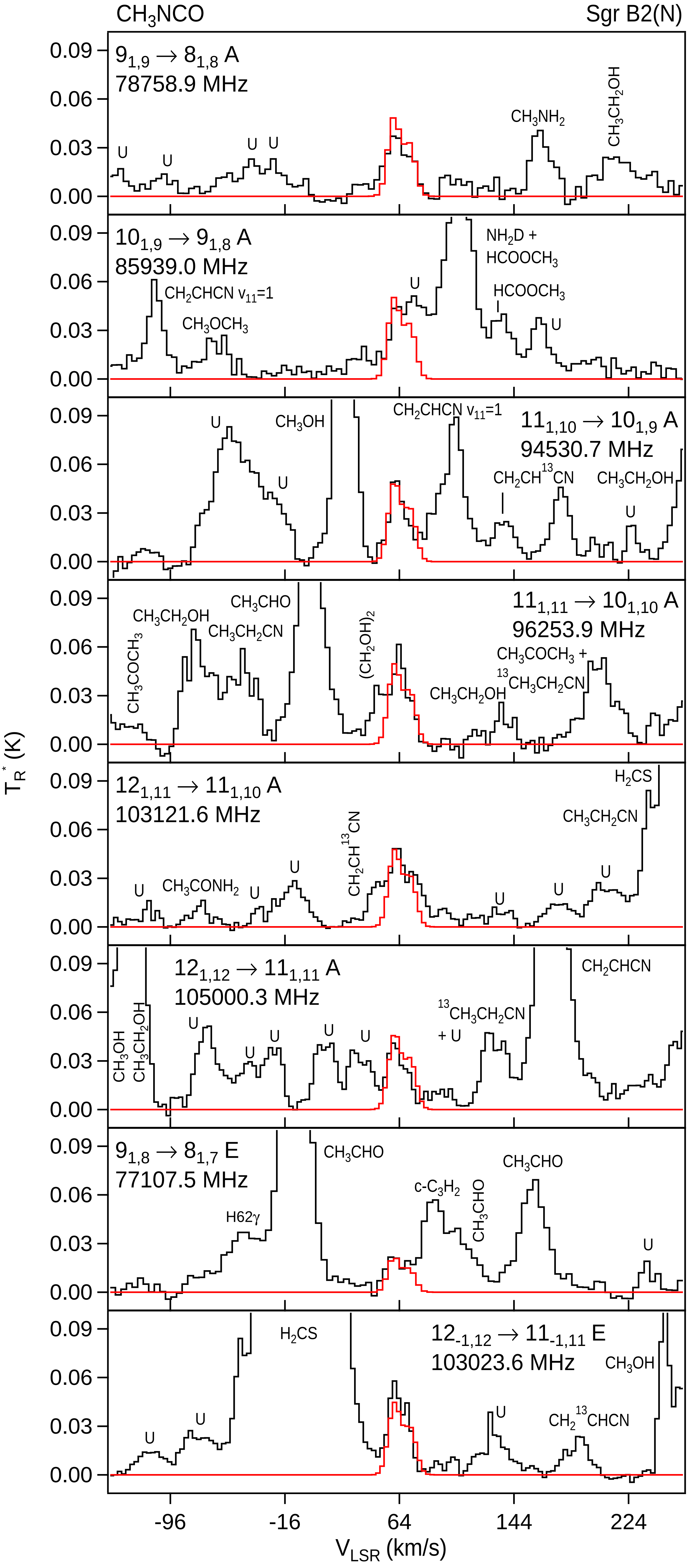

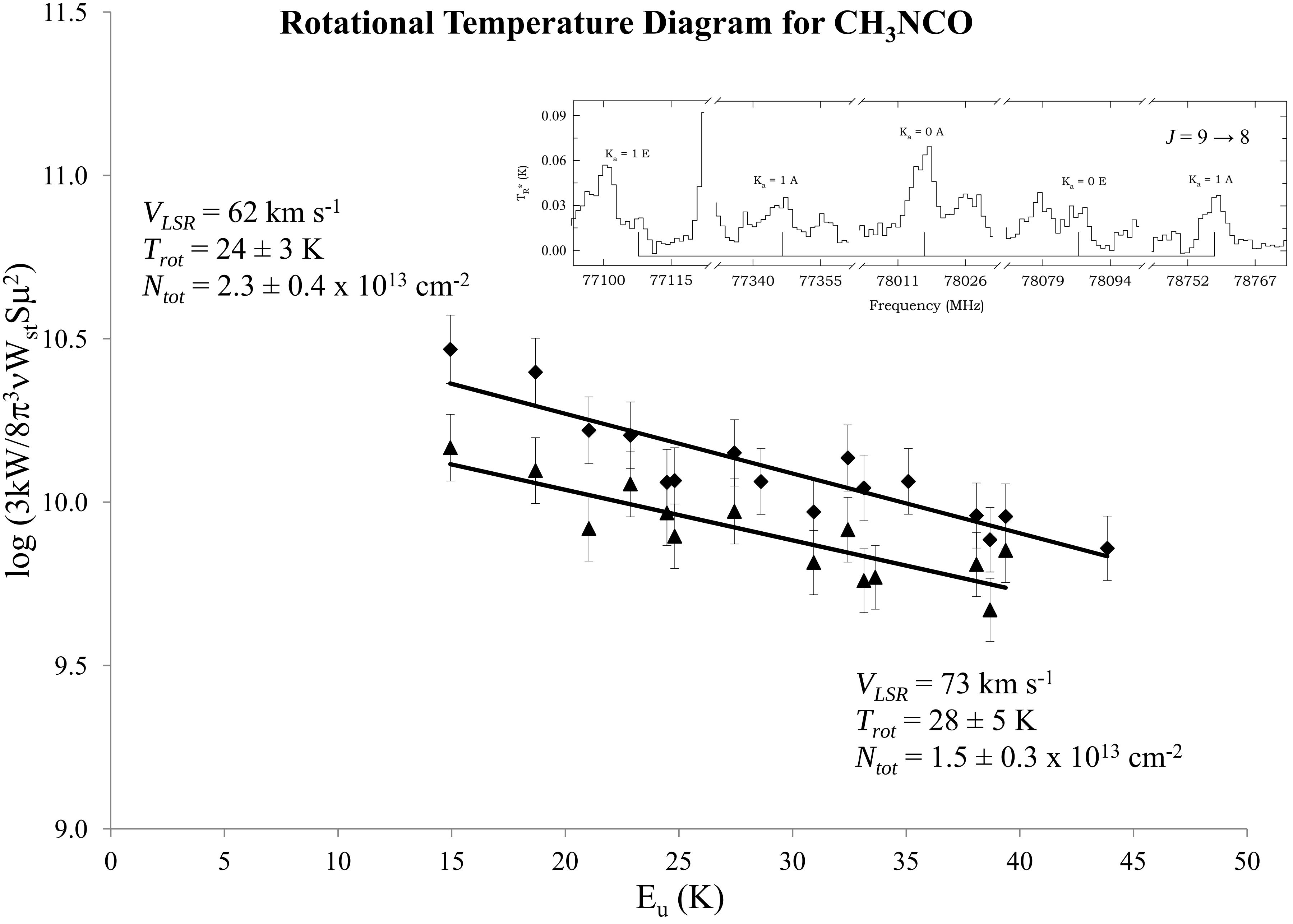

Rotational Temperature Diagram for CH3NCO
V_LSR = 62 km s-1
T_rot = 24 ± 3 K
N_tot = 2.3 ± 0.4 x 10^13 cm-2
V_LSR = 73 km s-1
T_rot = 28 ± 5 K
N_tot = 1.5 ± 0.3 x 10^13 cm-2
log (3kW/8π^3νW_stSμ^2)
E_u (K)
11.5
11.0
10.5
10.0
9.5
9.0
0
5
10
15
20
25
30
35
40
45
50
J = 9 → 8
Ka = 1 E
Ka = 1 A
Ka = 0 A
Ka = 0 E
Ka = 1 A
T_R* (K)
0.09
0.06
0.03
0.00
77100
77115
77340
77355
78011
78026
78079
78094
78752
78767
Frequency (MHz)

Title: Interstellar Detection of Methyl Isocyanate CH3NCO in Sgr B2(N):
A Link from Molecular Clouds to Comets
Authors: Halfen D.T., Ilyushin V.V., Ziurys L.M.
Table: Previous and New Rest Frequencies of CH3NCO (vb = 0) of the
Revised Analysis

================================================================================

Byte-by-byte Description of file: apjl520369t1_mrt.txt

--------------------------------------------------------------------------------

| Bytes | Format | Units | Label | Explanations |
|---|---|---|---|---|
| 1- 2 | I2 | --- | J' | Upper J quantum number |
| 4- 5 | I2 | --- | Ka' | Upper Ka quantum number |
| 7- 8 | I2 | --- | Kc' | Upper Kc quantum number |
| 10- 11 | I2 | --- | J" | Lower J quantum number |
| 13- 14 | I2 | --- | Ka" | Lower Ka quantum number |
| 16- 17 | I2 | --- | Kc" | Lower Kc quantum number |
| 19 | A1 | --- | A/E | A/E symmetry |
| 21- 22 | I2 | --- | m | Internal rotation quantum number, m |
| 24- 33 | F10.3 | MHz | Freq | Frequency of the transition |
| 35- 40 | F6.3 | MHz | O-C | Observed minus calculated value |
| 42 | A1 | --- | Source | Source of frequency (1) |

--------------------------------------------------------------------------------

Note (1): Below 40 GHz, frequencies from Koput (1986) [1986JMoSp.115..131K],
above from this work, unless specified:
a = From Kasten & Dreizler (1986) [1986ZNatA..41..637K].
b = Estimated from observed spectra in Figure 1.

--------------------------------------------------------------------------------

| | | | | | | | | | | |
|---|---|---|---|---|---|---|---|---|---|---|
| 1 | 0 | 1 | 0 | 0 | 0 | A | 0 | 8671.369 | 0.234 | a |
| 1 | 0 | 1 | 0 | 0 | 0 | E | -2 | 8712.440 | -0.006 | |
| 1 | 0 | 1 | 0 | 0 | 0 | E | 1 | 8714.600 | 0.248 | |
| 1 | 0 | 1 | 0 | 0 | 0 | A | 3 | 8722.070 | -0.037 | |
| 1 | 0 | 1 | 0 | 0 | 0 | A | -3 | 8722.930 | 0.289 | |
| 1 | 0 | 1 | 0 | 0 | 0 | E | 4 | 8737.510 | 0.172 | |
| 2 | 1 | 1 | 1 | 1 | 0 | E | 1 | 17131.350 | -0.306 | |
| 2 | -1 | 2 | 1 | -1 | 1 | E | 1 | 17139.540 | 0.925 | |
| 2 | 1 | 1 | 1 | 1 | 0 | A | 0 | 17190.371 | 0.602 | a |
| 2 | 0 | 2 | 1 | 0 | 1 | A | 0 | 17342.529 | 0.448 | a |
| 2 | 1 | 1 | 1 | 1 | 0 | E | -2 | 17404.820 | 0.224 | |
| 2 | -1 | 2 | 1 | -1 | 1 | E | -2 | 17413.270 | 0.214 | |
| 2 | 0 | 2 | 1 | 0 | 1 | E | -2 | 17424.730 | -0.135 | |
| 2 | 0 | 2 | 1 | 0 | 1 | E | 1 | 17426.090 | 0.535 | |
| 2 | 0 | 2 | 1 | 0 | 1 | A | 3 | 17444.100 | -0.122 | |
| 2 | 0 | 2 | 1 | 0 | 1 | A | -3 | 17445.790 | 0.498 | |
| 2 | 1 | 1 | 1 | 1 | 0 | A | 3 | 17448.710 | 0.001 | |
| 2 | 1 | 2 | 1 | 1 | 1 | A | 3 | 17454.140 | 0.133 | |
| 2 | 0 | 2 | 1 | 0 | 1 | E | 4 | 17474.880 | 0.218 | |
| 2 | -1 | 2 | 1 | -1 | 1 | E | 4 | 17491.620 | 0.172 | |
| 2 | 1 | 2 | 1 | 1 | 1 | A | 0 | 17504.688 | 0.539 | a |
| 2 | 1 | 2 | 1 | 1 | 1 | A | -3 | 17510.450 | 0.807 | |
| 3 | 2 | 1 | 2 | 2 | 0 | E | 1 | 25664.890 | -0.183 | |
| 3 | 1 | 2 | 2 | 1 | 1 | E | 1 | 25696.960 | -0.812 | |
| 3 | -1 | 3 | 2 | -1 | 2 | E | 1 | 25711.200 | 1.081 | |
| 3 | 1 | 2 | 2 | 1 | 1 | A | 0 | 25785.410 | 0.910 | |
| 3 | 0 | 3 | 2 | 0 | 2 | A | 0 | 26013.340 | 0.691 | |
| 3 | 2 | 1 | 2 | 2 | 0 | A | 0 | 26038.200 | -0.050 | |
| 3 | 1 | 2 | 2 | 1 | 1 | E | -2 | 26109.160 | 0.035 | |
| 3 | -1 | 3 | 2 | -1 | 2 | E | -2 | 26119.520 | -0.050 | |
| 3 | 0 | 3 | 2 | 0 | 2 | E | 1 | 26131.320 | 0.791 | |
| 3 | 0 | 3 | 2 | 0 | 2 | E | -2 | 26137.030 | -0.201 | |
| 3 | 0 | 3 | 2 | 0 | 2 | A | 3 | 26166.190 | -0.162 | |
| 3 | 0 | 3 | 2 | 0 | 2 | A | -3 | 26168.670 | 0.705 | |
| 3 | 1 | 2 | 2 | 1 | 1 | A | 3 | 26172.870 | -0.143 | |
| 3 | 1 | 3 | 2 | 1 | 2 | A | 3 | 26181.160 | 0.201 | |
| 3 | 0 | 3 | 2 | 0 | 2 | E | 4 | 26212.280 | 0.322 | |
| 3 | -1 | 3 | 2 | -1 | 2 | E | 4 | 26236.710 | -0.560 | |
| 3 | 1 | 2 | 2 | 1 | 1 | E | 4 | 26237.850 | -0.074 | |
| 3 | 1 | 3 | 2 | 1 | 2 | A | 0 | 26256.710 | 0.707 | |
| 3 | 1 | 2 | 2 | 1 | 1 | A | -3 | 26257.660 | 0.331 | |
| 3 | 1 | 3 | 2 | 1 | 2 | A | -3 | 26265.680 | 0.459 | |
| 4 | 2 | 2 | 3 | 2 | 1 | E | 1 | 34225.420 | 0.153 | |
| 4 | 1 | 3 | 3 | 1 | 2 | E | 1 | 34262.960 | -1.285 | |
| 4 | -1 | 4 | 3 | -1 | 3 | E | 1 | 34285.420 | 1.262 | |
| 4 | 1 | 3 | 3 | 1 | 2 | A | 0 | 34380.140 | 1.092 | |
| 4 | 3 | 1 | 3 | 3 | 0 | A | 0 | 34596.030 | -0.029 | |
| 4 | 3 | 2 | 3 | 3 | 1 | A | 0 | 34596.030 | -0.030 | |
| 4 | 0 | 4 | 3 | 0 | 3 | A | 0 | 34683.460 | 0.809 | |
| 4 | 2 | 3 | 3 | 2 | 2 | A | 0 | 34716.910 | 0.343 | |
| 4 | 2 | 2 | 3 | 2 | 1 | A | 0 | 34718.110 | 0.030 | |
| 4 | -3 | 2 | 3 | -3 | 1 | E | 1 | 34746.730 | -1.932 | |
| 4 | 1 | 3 | 3 | 1 | 2 | E | -2 | 34816.500 | 0.240 | |
| 4 | -2 | 3 | 3 | -2 | 2 | E | 1 | 34822.710 | -0.252 | |
| 4 | -1 | 4 | 3 | -1 | 3 | E | -2 | 34825.770 | -0.298 | |
| 4 | 0 | 4 | 3 | 0 | 3 | E | 1 | 34827.320 | 0.979 | |
| 4 | 0 | 4 | 3 | 0 | 3 | E | -2 | 34849.040 | -0.477 | |
| 4 | 0 | 4 | 3 | 0 | 3 | A | 3 | 34888.140 | -0.365 | |
| 4 | 0 | 4 | 3 | 0 | 3 | A | -3 | 34891.510 | 0.839 | |
| 4 | 1 | 3 | 3 | 1 | 2 | A | 3 | 34896.830 | -0.426 | |
| 4 | 1 | 4 | 3 | 1 | 3 | A | 3 | 34907.990 | 0.143 | |
| 4 | 0 | 4 | 3 | 0 | 3 | E | 4 | 34949.500 | 0.290 | |
| 4 | 3 | 1 | 3 | 3 | 0 | E | 1 | 34961.090 | 0.678 | |
| 4 | -1 | 4 | 3 | -1 | 3 | E | 4 | 34982.250 | -0.959 | |
| 4 | 1 | 3 | 3 | 1 | 2 | E | 4 | 34983.610 | -0.162 | |
| 4 | 1 | 4 | 3 | 1 | 3 | A | 0 | 35008.430 | 0.836 | |
| 4 | 1 | 3 | 3 | 1 | 2 | A | -3 | 35011.480 | 0.285 | |
| 4 | 1 | 4 | 3 | 1 | 3 | A | -3 | 35022.190 | 0.478 | |
| 7 | -1 | 7 | 6 | -1 | 6 | E | 1 | 60029.044 | 1.291 | |
| 7 | 1 | 6 | 6 | 1 | 5 | A | 0 | 60160.932 | -0.051 | |
| 7 | 0 | 7 | 6 | 0 | 6 | A | 0 | 60688.191 | 0.805 | |
| 7 | 0 | 7 | 6 | 0 | 6 | E | 1 | 60833.422 | -0.101 | |
| 7 | 1 | 7 | 6 | 1 | 6 | A | 0 | 61260.729 | 0.826 | |
| 8 | -1 | 8 | 7 | -1 | 7 | E | 1 | 68617.550 | -0.011 | |
| 8 | 1 | 7 | 7 | 1 | 6 | A | 0 | 68753.890 | -0.302 | |
| 8 | 0 | 8 | 7 | 0 | 7 | A | 0 | 69353.888 | 0.635 | |
| 8 | 0 | 8 | 7 | 0 | 7 | E | 1 | 69469.196 | -0.306 | |
| 8 | 1 | 8 | 7 | 1 | 7 | A | 0 | 70009.179 | -0.380 | |
| 9 | 1 | 8 | 8 | 1 | 7 | E | 1 | 77107.478 | 0.010 | |
| 9 | -1 | 9 | 8 | -1 | 8 | E | 1 | 77211.707 | -0.357 | |
| 9 | 1 | 8 | 8 | 1 | 7 | A | 0 | 77347.355 | 0.438 | |
| 9 | 0 | 9 | 8 | 0 | 8 | A | 0 | 78017.932 | 0.309 | |
| 9 | 0 | 9 | 8 | 0 | 8 | E | 1 | 78087.673 | 0.456 | |
| 9 | 1 | 9 | 8 | 1 | 8 | A | 0 | 78758.868 | 0.357 | |
| 10 | -1 | 10 | 9 | -1 | 9 | E | 1 | 85811.557 | 0.255 | |
| 10 | 1 | 9 | 9 | 1 | 8 | A | 0 | 85938.967 | -0.134 | |
| 10 | 0 | 10 | 9 | 0 | 9 | A | 0 | 86680.195 | -0.120 | |
| 10 | 0 | 10 | 9 | 0 | 9 | E | 1 | 86686.575 | -0.133 | |
| 10 | 1 | 10 | 9 | 1 | 9 | A | 0 | 87506.605 | -0.066 | |
| 11 | 1 | 10 | 10 | 1 | 9 | A | 0 | 94530.000 | -0.686 | b |
| 11 | 0 | 11 | 10 | 0 | 10 | E | 1 | 95269.000 | 0.208 | b |
| 11 | 0 | 11 | 10 | 0 | 10 | A | 0 | 95340.000 | -1.148 | b |
| 11 | 1 | 11 | 10 | 1 | 10 | A | 0 | 96253.000 | -0.950 | b |
| 12 | -1 | 12 | 11 | -1 | 11 | E | 1 | 103022.000 | -1.611 | b |

```
12  1 11 11  1 10 A  0 103120.000 -1.615 b
12  0 12 11  0 11 A  0 103998.000 -1.944 b
12  1 12 11  1 11 A  0 104999.000 -1.262 b
```